\documentclass{webofc}
\usepackage[varg]{txfonts}   
\newcommand{\be}{\begin{equation}}
\newcommand{\ee}{\end{equation}}

\def\Lx{\Lambda}
\def\Hc{{\cal H}}
\newcommand{\etat}{{\tilde{\eta}}}
\newcommand{\bk}{{\bf k}}
\newcommand{\rhoz}{\rho_m}
\newcommand{\dpp}{\delta p}
\newcommand{\drho}{\delta \rho}
\woctitle{QUARKS-2016}
\begin{document}
\title{Effective description of dark matter as a viscous fluid}
%
%

\author{
\firstname{Stefan} \lastname{Floerchinger}\inst{1}
\and
\firstname{Mathias} \lastname{Garny}\inst{2}
\and
\firstname{Nikolaos} \lastname{Tetradis}\inst{2,3}\fnsep\thanks{\email{ntetrad@phys.uoa.gr}} 
\and
\firstname{Urs Achim} \lastname{Wiedemann}\inst{2}
}

\institute{
		Institut f\"{u}r Theoretische Physik, Universit\"{a}t Heidelberg, 69120 Heidelberg, Germany
\and 
          Theoretical Physics Department, CERN, CH-1211 Gen\`eve 23, Switzerland
          \and
          Department of Physics, University of Athens, Zographou 157 84, Greece        
          }

\abstract{%
Treating dark matter at 
large scales as an effectively viscous fluid  
provides an improved framework for the calculation of the density and velocity
power spectra compared to the standard assumption of an ideal pressureless fluid. 
We discuss how this framework can be made concrete through an appropriate
coarse-graining procedure.
We also review results that demonstrate that it improves the
convergence of cosmological perturbation theory.
}
\maketitle
\section{Introduction}

The calculation of the power spectrum of cosmological fluctuations 
for stochastic initial conditions is an important and challenging task. 
It provides the basis for analyzing data on the large-scale structure of
the universe and can lead to constraints on the parameters of 
the cosmological model. The spectrum can be computed through 
$N$-body simulations. However, these are CPU-intensive, while they 
do not provide
an intuitive understanding of the underlying physical processes. 
The alternative option is an analytical treatment, 
which amounts to solving the collisionless Vlasov-Poisson
equation perturbatively in the density contrast
for appropriate classes of initial conditions \cite{bernardeau0}.
The use of perturbative techniques in this context is complicated by the
growth of non-linearities at short length scales. 
Within the strongly non-linear regime, velocity dispersion
and virialization becomes important. 

The inability of analytical methods to describe the short scales reliably suggests
to reformulate cosmological perturbation theory
based on an effective description, applicable only above 
some length scale.
The description should contain effective
parameters which 
absorb the effect of short-scale perturbations that are `integrated out'~\cite{baumann,effective,Pietroni:2011iz}. 
In the conventional description of dark matter
as a pressureless ideal fluid, only the lowest moments 
of the phase-space distribution are taken into account. 
Enlarging this framework by including higher moments results
in viscous transport coefficients that parametrize a non-ideal stress tensor
for the dark matter. 
In the following we present heuristic arguments
that support the suggestion that an efficient fluid description of dark matter
must allow for such non-ideal terms. The role of the effective viscosity and sound velocity is to account for 
the interaction of large-scale fluctuations (with
wavenumbers $k < k_m$) with the 
short-scale ones ($k > k_m$) that are not followed explicitly 
in a fluid description limited to $k<k_m$.
In our analysis the UV contributions are incorporated in the large-scale theory 
at the level of a one-loop approximation. 
Going beyond one-loop can be achieved through 
renormalization-group techniques \cite{Max1,floerchinger}.
The approximation that we review here provides an intuitive and rather simple framework, while it still
resolves the
main deficiency of standard perturbation theory (SPT), namely 
the strong dependence on the short-scale dynamics that are out of the reach of
analytical techniques. 
We review the calculation of the density power 
spectrum presented in ref. \cite{blas} in the context of the coarse-grained
theory. 
We also present results for the velocity and cross spectra within
the same framework. 
 
\section{The relevant scales}

Our approach to the problem of cosmological matter perturbations assumes
the presence of two scales: the momentum scale $k_\Lx$ at which a fluid
description becomes feasible, and the scale $k_m < k_\Lx$ at which the 
description can be based on a small number of 
effective parameters. In practice we
can take $k_\Lx \sim 1-3 ~h/$Mpc, corresponding to length scales 
$\sim 3-10$ Mpc, and $k_m\sim 0.5-1~h/$Mpc, corresponding to length scales
$\sim 10-20$ Mpc. At the scale $k_m$ the density contrast is of order 1, while
above the scale $k_\Lx$ it is much larger than 1 and the
dynamics involves shell crossing, making a fluid description unfeasible.
The description at $k_m$ must include viscosity terms arising either
through the coarse-graining of the scales $k>k_m$ (effective viscosity), or
fundamental dark matter interactions (fundamental viscosity). 
The scales $k>k_\Lx$ correspond to virialized structures, which are
expected to have a negligible effect on long-distance dynamics \cite{Peebles:1980}. It seems 
reasonable then to expect that the effective viscosity results mainly  
from the integration of the modes with $k_m<k<k_\Lx$. The form of the linear
power spectrum (which falls off roughly as $k^{-3}$) is consistent 
with this expectation: the
effective viscosity at $k_m$ is dominated by the modes with $k$ slightly 
above this value, while the scales $k>k_\Lx$ give a negligible contribution. 
This efficient decoupling of deep UV modes is also supported by numerical
evidence from $N$-body simulations as well as analytical arguments based on
non-perturbative relations of response functions \cite{nishimichi,Garny:2015oya}.

There are several reasons for introducing the scale $k_m$ 
and distinguishing it from $k_\Lx$:
\begin{itemize}
\item
In the absence of dark-matter interactions,
the form of the effective terms beyond the perfect-fluid
description (such as effective viscosity) at the scale $k_m$ is in a certain sense universal. 
They do not depend strongly
on the initial conditions at $k_\Lx$, but only on $k_m$. 
If fundamental interactions exist, the fundamental viscosity should
vary only little between $k_\Lx$ and $k_m$, so that it
can be estimated at $k_m$ without loss of accuracy. 
\item
The main role of the scale $k_m$ is to act as an UV cutoff for perturbative
corrections in the large-scale theory, thus eliminating 
contributions of dubious validity, arising from the modes near and above $k_\Lx$. This effect is seen clearly in the good convergence of our results, 
in contrast to standard perturbation theory \cite{blas}. 
\item
The scale $k_m$ can be lowered even further than the range that we are
considering, if one would like to focus on the effective theory for length scales 
much larger than the range of baryon acoustic oscillations (BAO). 
In such a case, the renormalization-group improvement of the 
effective description is crucial \cite{floerchinger}. 
\end{itemize}

\section{Estimate of the effective viscosity}

The validity of the fluid description for cold dark matter cannot be established
similarly to systems close to thermal equilibrium. The crucial elements supporting this description are the small dark-matter velocity and
the finite age of the universe, during which dark matter particles can
drift over a finite distance, much smaller than the Hubble radius.
A more quantitative discussion of this point is given in ref. \cite{baumann}.
In the absence of dark-matter interactions, one starts from the Vlasov-Poisson 
equations 
and writes the particle phase-space density as
\be
f({\bf x},{\bf p},\tau)=f_0(p)[1+\delta_f({\bf x},p,\hat{\bf p},\tau)],
\label{phase} \ee
with $\tau$ the conformal time.
The Fourier modes of the perturbation are then expanded in terms of 
Legendre polynomials $P_n$:
\be
\delta_f({\bf k},p,\hat{\bf p},\tau)=
\sum_{n=0}^\infty (-i)^n (2n+1)\,\delta_f^{[n]}({\bf k},p,\tau)
\, P_n(\hat{\bf k}\cdot \hat{\bf p}),
\label{legendre} \ee
with unit vectors $\hat{\bf k}$, $\hat{\bf p}$.
The Vlasov-Boltzmann equation implies a hierarchy of evolution equations for the 
higher moments:
\be
\frac{d\delta_f^{[n]}}{d\tau}=k v_p \left[\frac{n+1}{2n+1}\delta_f^{[n+1]}
-\frac{n}{2n+1}\delta_f^{[n-1]} \right], 
~~~~~~~~~n\geq2,
\label{momev} \ee
with $v_p=p/{am}$ the particle velocity.
If the typical value of 
$v_p$ is sufficiently small, the higher moments can be neglected.

A quantitative constraint can be obtained if one assumes that the 
particle velocity
is of the order of the fluid velocity $v$ at small length scales. 
We emphasize that this assumption is valid only for cold dark matter, 
for which the particle velocity is induced by the gravitational fields 
generated by the growth of large-scale structure. 
The assumption is supported by the estimated typical velocities 
of dark matter particles in galaxy haloes.\footnote{Due to the cosmological expansion, the massive particles 
that constitute the cold dark matter have negligible velocity 
originating in their thermal motion before freeze-out. 
Warm dark-matter particles would have much larger velocities, which
makes the application of the fluid description to this case problematic.}
At the comoving momentum scale $k$ we expect $[\theta/\Hc]^2\sim k^3 P^L(k)$,
with $P^L(k)$ the linear power spectrum evaluated on the growing mode, 
$\theta=\vec{\bf k}\cdot \vec{\bf v}$ the velocity divergence, and $\Hc=(1/a)(da/d\tau)$.
The linear spectrum scales roughly as $k^{-3}$ above a scale $\sim k_m$,
so that $k^3 P^L(k)$ is roughly constant, with a value of order 1 today.
Its time dependence is given by $D_L^2$,
 with $D_L$ the linear growth factor. 
 We expect then that the maximal
particle velocity can be identified with the fluid velocity at the 
scale $k_m$, and is roughly 
\be
v_p\sim \frac{\Hc}{k_m}\, D_L. 
\label{partvel} \ee
The time $\tau$ 
available for the higher moments to grow
is $\sim 1/\Hc$. 
Therefore, the dimensionless factor characterizing the
growth of higher moments is $k v_p/\Hc\sim D_L\, k/k_m$.  
Scales with $k\gg k_m/D_L$ require the use 
of the whole Boltzmann hierarchy and cannot be treated through a fluid
description that takes into account only the lowest moments. 
In practice, the validity 
of the fluid description extends beyond $k_m$ at all times, 
as the initial values of the
higher moments are much smaller than 1, while
$D_L\leq 1$. 
Therefore, on large scales, cold dark matter can be described as a cosmological 
fluid, not because microscopic interaction rates are sufficiently large to maintain local thermal equilibrium, but because the lifetime 
of the universe is too short for dark matter 
to deviate strongly from local equilibrium. 

The above picture can lead to an estimate of the effective viscosity 
$\eta$ that
must be attributed to
the dark-matter fluid because of the particle velocities 
that deviate from the uniform motion. 
In general, we expect $\eta/(\rho+p)\sim l_{\rm free}v_p$, with 
$l_{\rm free}$ the mean free path. For the effective viscosity we 
can estimate $l_{\rm free}\sim v_p /H$, with $H=\Hc/a$. 
In this way we obtain for the kinematic viscosity of the 
dark-matter fluid
\be
\nu_{\rm eff} \Hc =\frac{\eta_{\rm eff}}{\rho_m \, a}\Hc
\sim l_{\rm free}v_p H\sim \frac{\Hc^2}{k_m^2} D^2_L,
\label{etaeff} \ee
where we have neglected the pressure.
The numerical value is small, of the order of $10^{-6}$ today,
and even smaller at higher redshifts. However, viscosity plays a significant role in the context of the coarse-grained theory.

We can also estimate the viscosity induced by possible fundamental interactions between the dark-matter particles.
For interacting dark matter, with number density $n$, mass $m$ and
cross section $\sigma$, the mean free path is $l_{\rm free}\sim 1/(n\sigma)$. 
We expect then fundamental viscosity 
\be
\nu_{\rm fund} \Hc =\frac{\eta_{\rm fund}}{\rho_m \, a}\Hc
\sim l_{\rm free}v_p H \sim \frac{1}{n\sigma} \frac{H^2}{k_m}a D_L 
\sim \frac{m}{\sigma} \frac{8\pi G_N}{3k_m} \frac{a D_L }{\Omega_m}.
\label{etafund} \ee
The time dependence, given by the last factor, indicates that the fundamental
viscosity drops very quickly at early times.
We point out that the above estimate is valid only for cold dark matter, 
for which the typical particle velocity is induced by the late-time
growth of structure. For warm dark matter, the typical velocity 
and the resulting viscosity would evolve differently with time.

We note that that the fundamental viscosity of cold dark matter does not
grow indefinitely for decreasing $\sigma$. The mean free path
of dark matter particles cannot exceed $v_p/ H\sim a D_L /k_m$. 
As a result, for 
values of $\sigma$ such that $1/(n\sigma)$ exceeds $a D_L /k_m$,
the particle scattering has no effect.
This implies that the fundamental viscosity 
cannot take values larger than roughly $(\Hc^2/k_m^2) D_L^2$, and
is always subleading or comparable to the effective viscosity.

As a final remark, we would like to emphasize that the constraints 
derived in refs. \cite{skordis1,kunz,skordis2} for the non-ideal properties
of dark matter are not applicable to the case of cold dark matter that we
are considering. In these works the viscosity and sound velocity are
assumed to stay constant or even grow with redshift, in contrast to 
the redshift-dependence indicated by our eqs. (\ref{etaeff}) and (\ref{etafund}).
In our picture, the particle velocity is induced by the growth of structure
and is a late-time phenomenon. The assumptions of refs. \cite{skordis1,kunz,skordis2} can by justified only within a warm-dark-matter
scenario, in which the particles have velocities originating in their
primordial motion.

\section{One-loop determination of the effective viscosity}

For the construction of the coarse-grained theory we follow the 
Wilsonian approach. A detailed description, including a discussion of the
 renormalization-group flow of the coarse-grained theory, is given
in ref. \cite{floerchinger}. The main idea is that the
effective parameters of the large-scale theory result from the 
integration of the UV modes. For the problem at hand, the effective 
viscosity and sound velocity of the 
dark-matter fluid are given by one-loop expressions that depend on
the dimensionful scale 
\be
\sigma^2_{dm}({\etat})=\frac{4\pi}{3} \int_{k_m}^\infty dq\, P^L(q,\etat)
=\frac{4\pi}{3}  D^2_L(\etat) \int_{k_m}^\infty dq\, P^L(q,0),
\label{sigmadm} \ee
with $\etat=\ln a=-\ln(1+z)$ and $D_L(\etat)$ the linear growth factor:
$\delta(\bk,\etat)=D_L(\etat)\,\delta(\bk,0)$ on the growing mode.
For a spectrum that scales as $1/k^3$, the integral is dominated by the region near $k_m$.

A precise definition of the effective couplings can be given through 
 the propagator of theory. 
One first defines the doublet
\begin{equation}
 \left(
\begin{array}{c}
\varphi_{1}(\textbf{k},\etat)\\ \\ \varphi_{2}(\textbf{k},\etat)
\end{array}
\right)
=\left(
\begin{array}{c}
\delta_\bk(\etat)\\ \\-\dfrac{\theta_{\bk}( \etat)}{\mathcal{H}}
\end{array}
\right),
\label{oneloopprop}
\end{equation}
and the propagator 
\be
G_{ab}(\bk, \tilde\eta, \tilde\eta^\prime)\, \delta^{(3)}(\bk - \bk^\prime) = \left\langle  \frac{\delta \varphi_a(\bk,\tilde\eta)}{\delta \varphi_b(\bk^\prime, \tilde\eta^\prime)} \right\rangle\, .
\label{propagator}
\ee
The power spectrum is defined as 
\be
 \langle\varphi_{a}(\bk, \etat)\varphi_{b}(\bk', \etat)\rangle
= \delta^{(3)}\,({\bk+\bk'}) \, P_{ab}({\bf{k}}, \etat).
\label{spectrum}
\ee

We concentrate on an Einstein-de Sitter (EdS) background, for which
$D_L(\tilde\eta)=\exp(\tilde\eta)$ and the linear propagator for the growing mode in
the perfect-fluid theory is
\be
g_{ab}(\tilde\eta-\tilde\eta^\prime) = \frac{e^{\tilde\eta-\tilde\eta^\prime}}{5} \begin{pmatrix} 3 && 2 \\ 3 && 2 \end{pmatrix}.
\label{linprop} \ee
The {\it one-loop} contribution to the low-energy 
propagator with wavenumber $k \to 0$, coming from modes with $q\geq k_m$, is given by
\be
\delta g_{ab}(\bk, \tilde\eta,\tilde\eta^\prime) 
=  - k^2 e^{\tilde\eta-\tilde\eta^\prime} \sigma_{dm}^2(\tilde\eta) 
\begin{pmatrix} \frac{61}{350} && \frac{61}{525} \\ \frac{27}{50} && \frac{9}{25} \end{pmatrix}\,
\simeq - k^2 \, e^{\eta} \, \sigma_{dm}^2(\etat) \frac{3}{50}
\begin{pmatrix}3 \cdot 0.968 && 2\cdot 0.968 \\ 9 && 6 \end{pmatrix}\, ,
\label{onlooppro}
\ee
where we have taken the limit $\etat\gg \etat'$ for the second equality.
On the other hand, the {\it linear} propagator of a theory with small kinematic 
viscosity and sound velocity of the form
\be
\nu\Hc=\frac{\eta}{\rhoz\, a}\Hc=\frac{3}{4}\, \beta_\nu \, e^{2 \etat}\, \frac{\Hc^2}{k^2_m}\, ,
~~~~~~~~~~
c^2_s = \frac{\dpp}{\drho}= \beta_s\,  e^{2 \etat}\, \frac{\Hc^2}{k^2_m}\, ,
\label{deff} \ee
contains a contribution 
\be
\delta g_{ab}(\bk,\etat) = -\frac{k^2}{k^2_m} (\beta_\nu+\beta_s) 
\, \frac{e^{3 \etat}}{45} \,
\begin{pmatrix} 3 && 2 \\ 9 && 6 \end{pmatrix}
\label{corrprop} \ee
in addition to the perfect-fluid term (\ref{linprop}),
where we have kept only the leading contribution for $\etat\gg\etat'$.

We can now derive precise expressions for the effective parameters of the
large-scale theory, by identifying the 
{\it linear} contribution to the propagator (\ref{corrprop}) 
for the {\it viscous} theory, 
with the {\it one-loop} correction (\ref{onlooppro}) of the {\it perfect-fluid} propagator. This can be achieved with 1\% accuracy, and we 
find 
\be
\beta_s+\beta_\nu=\frac{27}{10} \, k^2_m\sigma^2_{dm}(0).
\label{matching} \ee
It must be emphasized that there are 
no free parameters in our approach, as the effective couplings are uniquely 
determined. Moreover, the deep UV region, which is out of the reach of 
perturbation theory, gives a negligible contribution to the integral of eq. 
(\ref{sigmadm}), consistently with the expectation of decoupling of virialized
structures. The dominant contribution comes from the region slightly above 
$k_m$, for which perturbation theory is expected to give reliable results. 

By matching heuristically viscous fluid dynamics to
the one-loop contribution (\ref{corrprop}), effective couplings were explored in \cite{blas}. This procedure
can be obtained as a limiting case from the renormalization-group formulation of Ref.~\cite{floerchinger}.
The latter approach also allows one to break the degeneracy between $\beta_\nu$ and $\beta_s$~\cite{floerchinger}. 
The numerical results that we present below depend only on the sum $\beta_\nu+\beta_s$ to a very good approximation.


\begin{figure*}
\centering
\includegraphics[width=6cm]{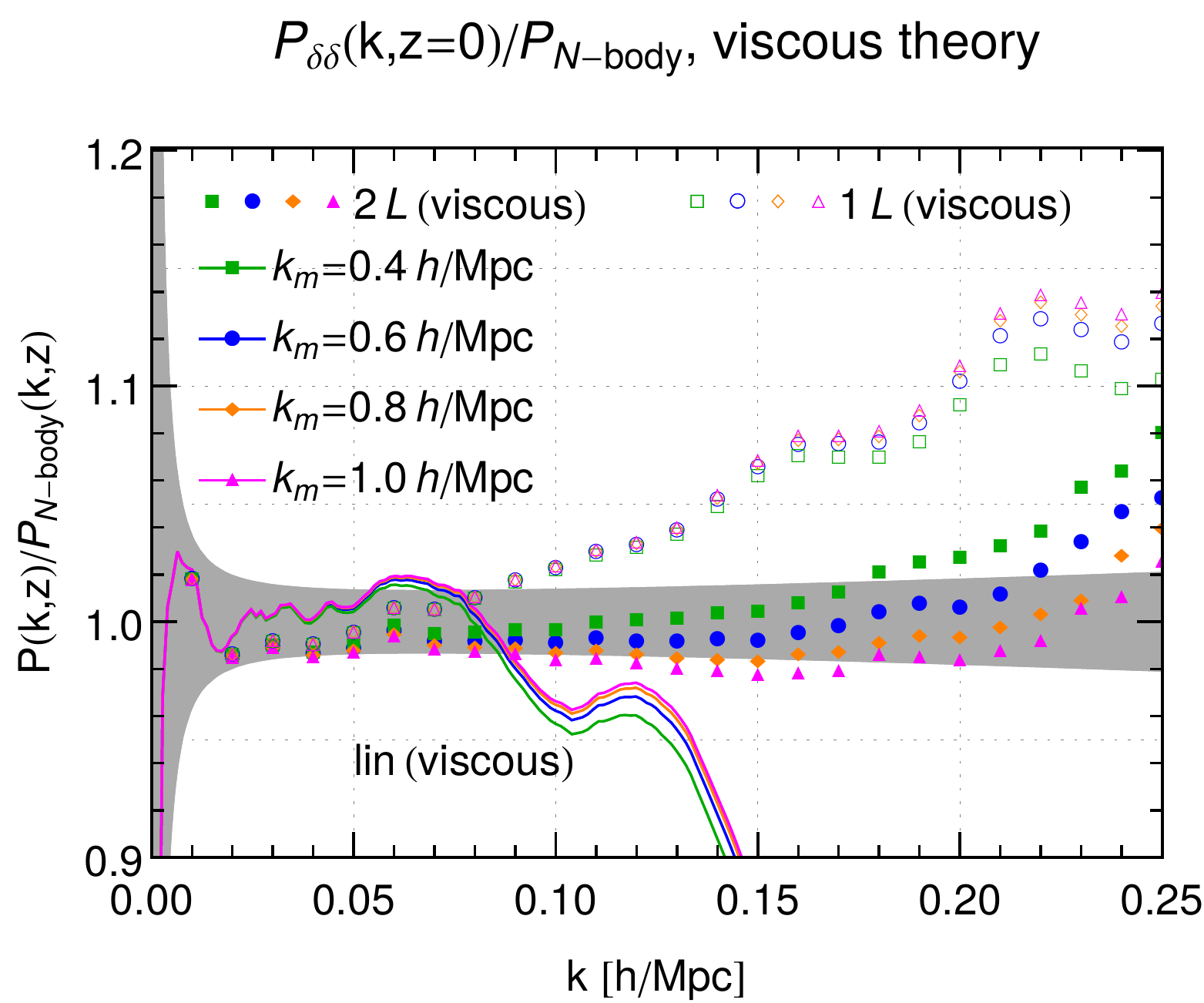}\qquad 
\includegraphics[width=6cm]{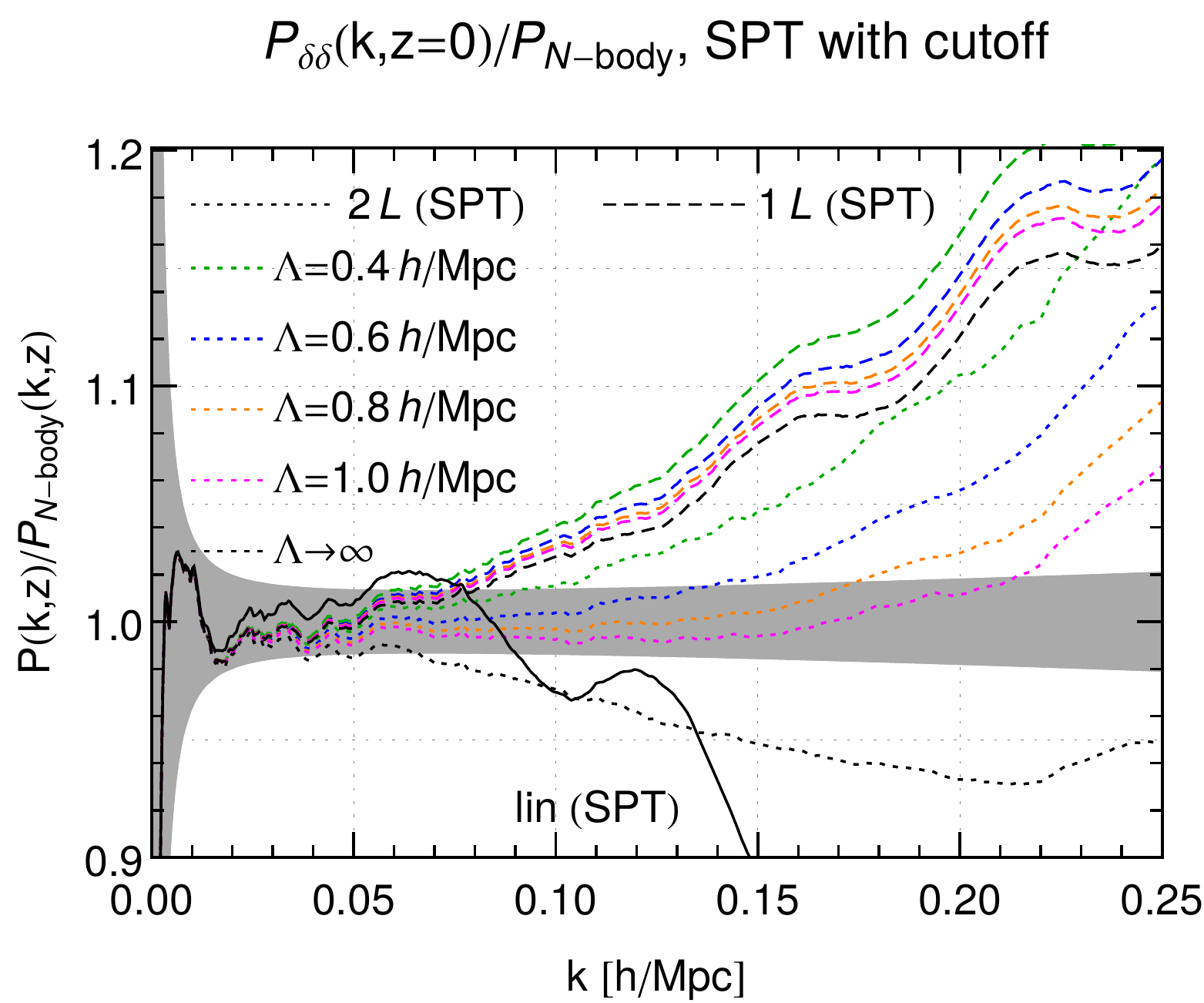}
\caption{Density power spectrum at redshift $z=0$,
normalized to the $N$-body power spectrum~\cite{Kim:2011ab}. The left panel 
shows the results obtained in the viscous
description, for various values of the scale 
$k_m$. 
The coloured lines show the linear spectra,  
the open and filled symbols the one- and two-loop
spectra.
For comparison, the right panel 
shows the corresponding one- and two-loop results
in standard perturbation theory. Here $\Lambda$ is a sharp cutoff in $k$-space. The grey shaded region shows the estimated uncertainty of the $N$-body simulation. Figure taken from Ref.~\cite{blas}. } 
\label{fig-2}      
\end{figure*}

\section{Density and velocity power spectra}

The procedure that we outlined in the previous section for an EdS background 
can be generalized to the $\Lambda$CDM case with an appropriate change of 
variables \cite{blas}.
We present here results for the density, velocity and mixed power spectra for 
a large-scale viscous theory with parameters determined through 
eq. (\ref{matching}).
The non-linear spectra for the low-energy theory can be computed
through perturbation theory, with the crucial modification that (i) all 
momentum integrals possess a UV cut off $k_m$, and (ii) the Euler equation
includes pressure and viscosity contributions. The presence of the cutoff
prevents the inclusion of spurious effects from the deep UV, so that 
fast convergence is expected. It must be emphasized that the precise value 
of the scale $k_m$ should not influence the final results for physical 
quantities, such as the power spectra: the explicit dependence of the effective
pressure and viscosity on $k_m$ compensates the $k_m$-dependence introduced
by the cutoff. In some sense, 
the scale $k_m$ may be viewed as
a technical device that guarantees the good convergence of our scheme. 
Nevertheless, because of the perturbative truncation we expect a residual dependence
that is a measure of the remaining theoretical uncertainty.

In order to test the accuracy of our results we compare them
to the power spectrum extracted from $N$-body simulations \cite{Kim:2011ab}.
In the left panel of fig.\,\ref{fig-2} 
we show the density power spectrum at redshift $z=0$,
obtained from solving the viscous fluid equations
and averaging over an initially Gaussian random distribution of the density field.
The different symbols show our results obtained for the four values
$k_m=0.4, 0.6, 0.8, 1.0\, h/$Mpc. In addition, open symbols are the one-loop results,
and the filled symbols show the power spectrum up to two-loops. The loop expansion is
perturbative in the density contrast. However, the effect of the time-dependent
pressure and viscosity on the propagation of density and velocity perturbations is
taken into account non-perturbatively. 
We normalized all curves to the result of the $N$-body simulation.
The grey-shaded area corresponds to an estimate of the error of the simulation.
The agreement between $N$-body data and the two-loop results is at the percent level for $k\lesssim 0.2\, h/$Mpc.
In addition, the variation of the results when changing the scale $k_m$ is mild, at the $\pm 2 \%$ level for
$k\lesssim 0.2\, h/$Mpc. Our results can be compared to the ones obtained in
standard perturbation theory (SPT) \cite{bernardeau0}, depicted in the
right panel of \ref{fig-2}.  For SPT we show the one- and two-loop contributions obtained when imposing a
cutoff $\Lambda$ in $k$-space. The cutoff dependence in SPT is sizable at two loops, and amounts to $\pm 5\%\ (\pm 10\%)$
for $0.4\, h/$Mpc$<\Lambda<1\, h/$Mpc ($0.4\, h/$Mpc$<\Lambda\lesssim 5\, h/$Mpc). 
It is apparent that, as compared to SPT, the viscous
description reduces the uncertainty generated by 
the treatment of UV modes significantly (left panel in fig.\,\ref{fig-2}).
The remaining uncertainty within the viscous framework is consistent with the expected impact of effects not captured by the fluid dynamical
description employed here \cite{Pueblas:2008uv}.

\begin{figure*}
\centering
\includegraphics[width=6cm]{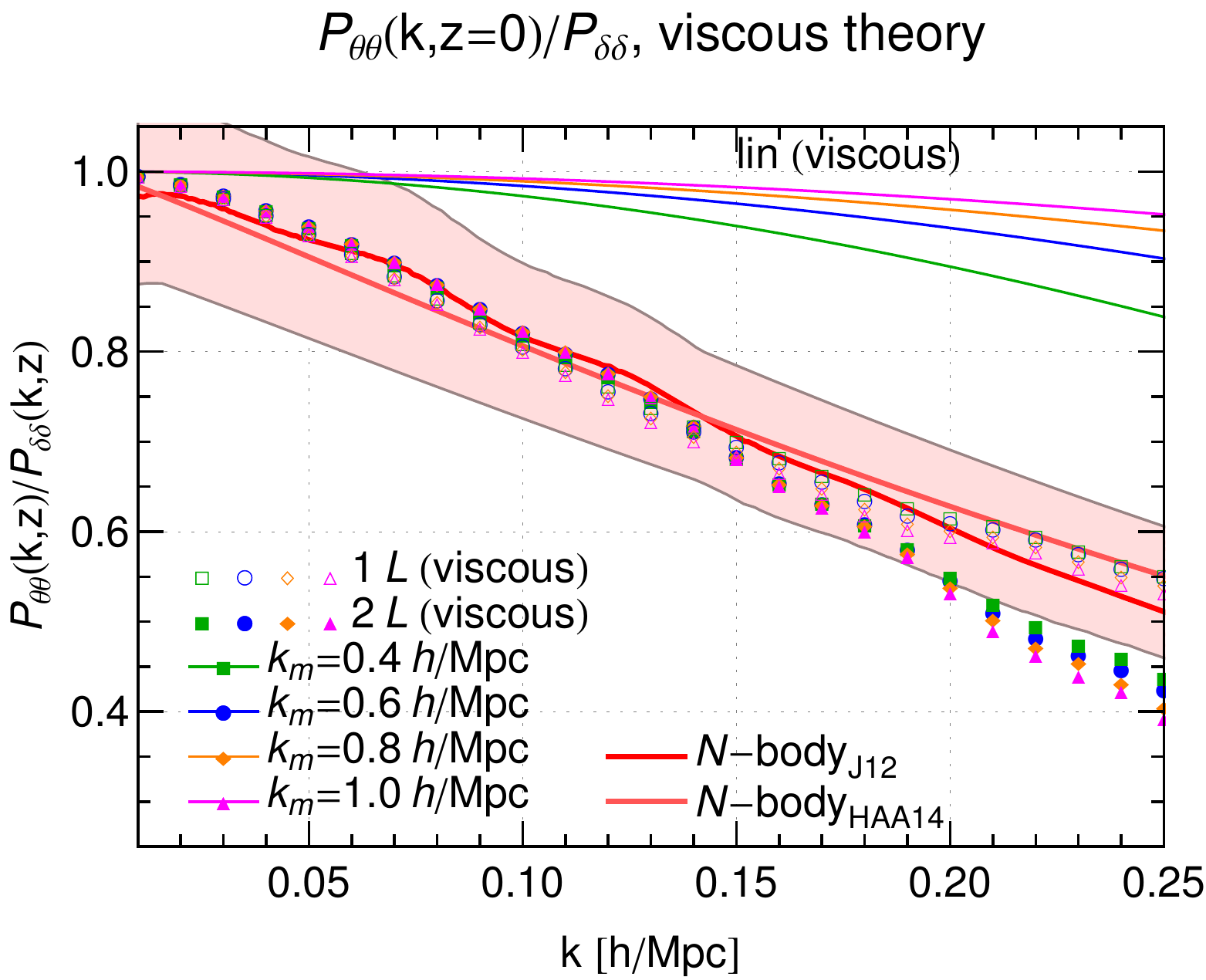}\qquad 
\includegraphics[width=6cm]{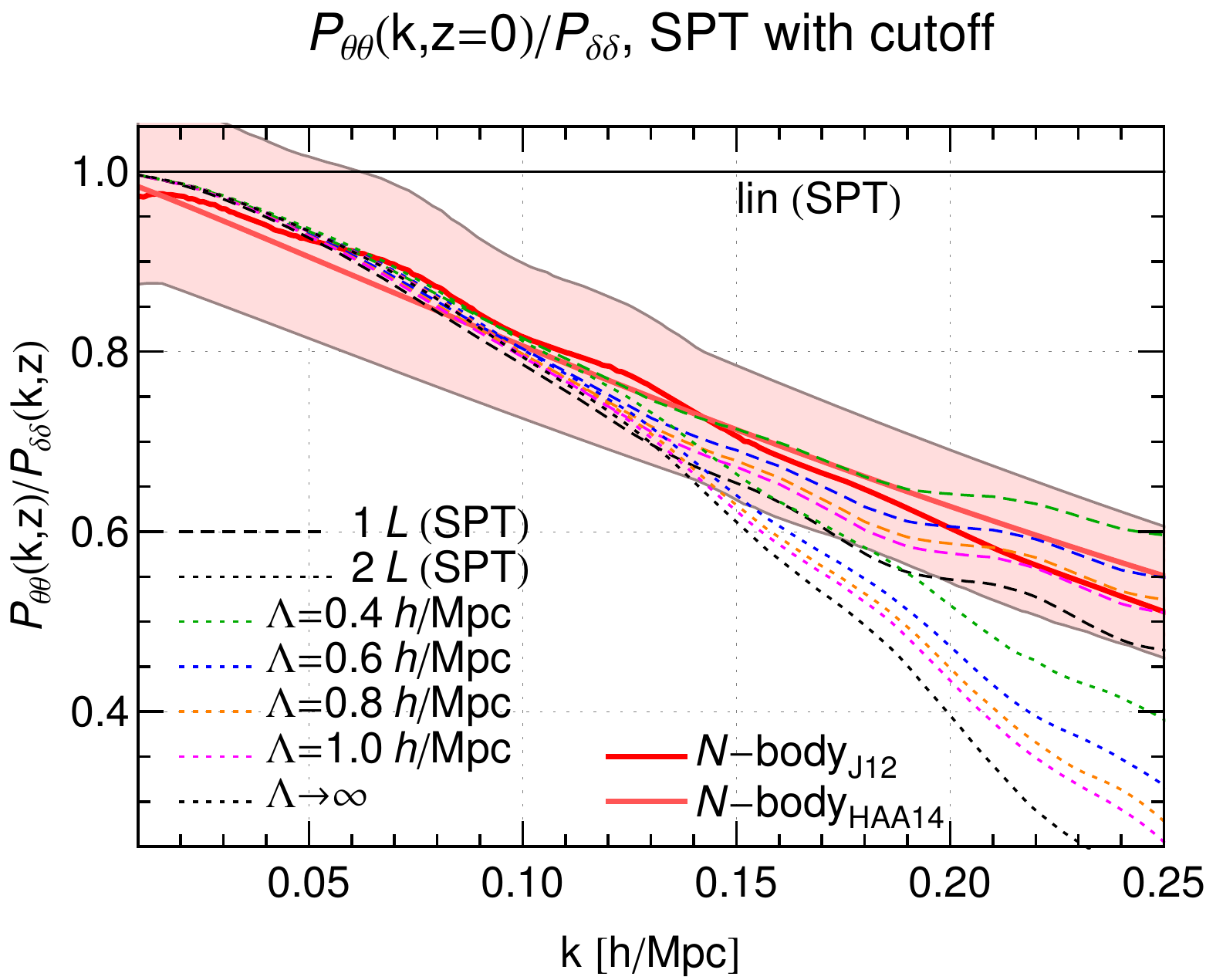}
\caption{Power spectrum of the velocity-divergence, normalized to the density power spectrum, for the viscous theory (left)
and standard perturbation theory (right). The various lines show the dependence on $k_m$ and on $\Lambda$, respectively, as
in fig.\,\ref{fig-2}. The red lines are $N$-body results ((J12) \cite{Jennings:2012ej}, (HAA14) \cite{Hahn:2014lca}), and the shaded
region is the quoted uncertainty of the velocity power spectra extracted from $N$-body data.
}
\label{fig-3}      
\end{figure*}

\begin{figure*}
\centering
\includegraphics[width=6cm]{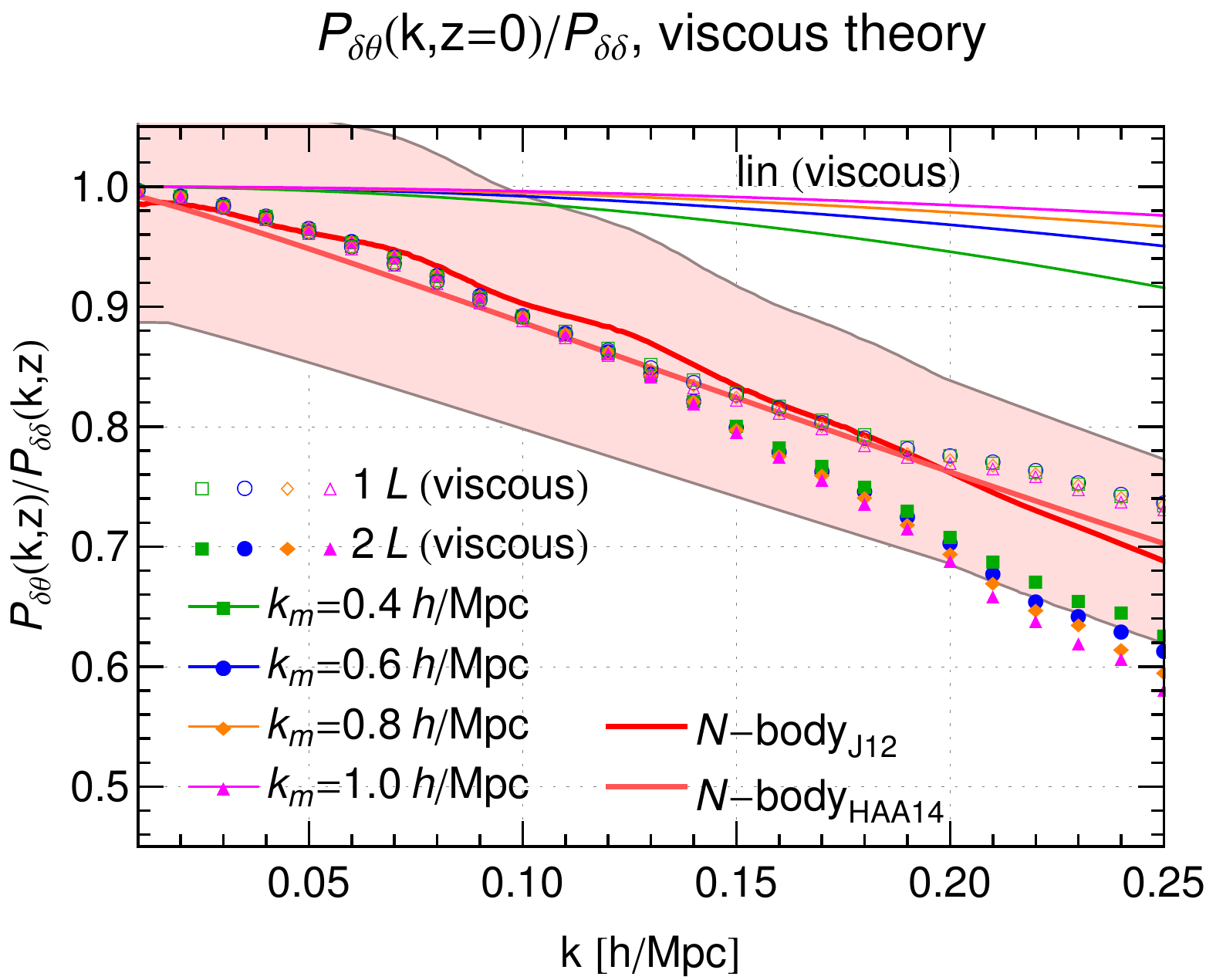}\qquad 
\includegraphics[width=6cm]{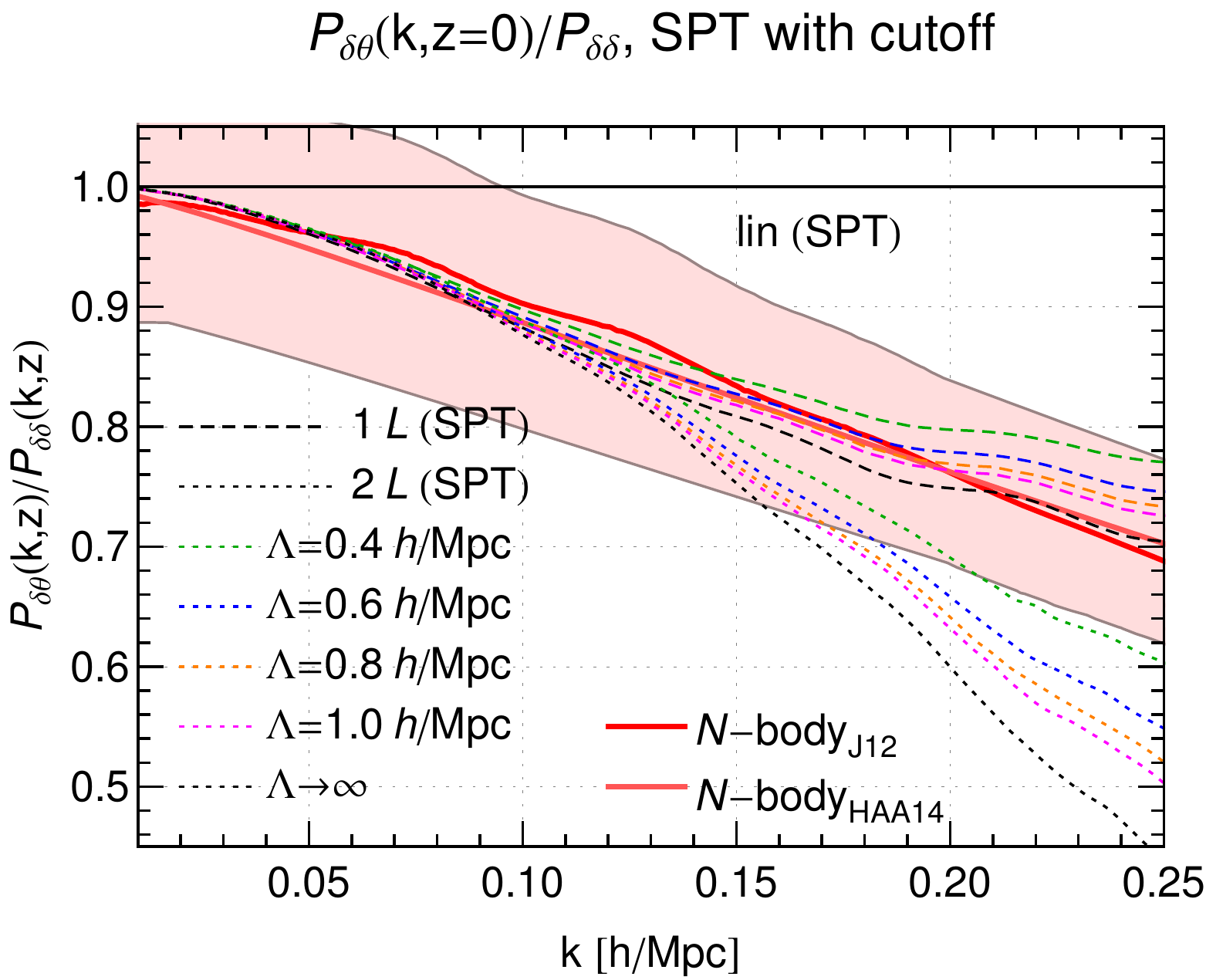}
\caption{Cross density-velocity power spectrum, 
corresponding to the same
approximations as in fig.\,\ref{fig-3}.}
\label{fig-4}      
\end{figure*}

We also consider 
the power spectrum for the velocity divergence and the cross power spectrum.
The velocity spectrum at $z=0$ is shown in fig.\,\ref{fig-3}, where we again compare the viscous description (left panel) with the result obtained
in SPT (right panel) for various values of $k_m$ and $\Lambda$, respectively. 
All curves are normalized to the density power spectra corresponding to the same approximation.
Fig.\,\ref{fig-4} shows a similar comparison for the cross power spectrum.
The extraction of power spectra for the velocity divergence from $N$-body simulations is less robust than
for the density, because of complications arising when extracting a continuous velocity field from a
discrete set of particles. We show the results obtained from two different methods \cite{Jennings:2012ej, Hahn:2014lca} by the two thick red lines
in figs.\,\ref{fig-3} and \ref{fig-4}. The results are sensitive to certain parameters of the simulation, especially the resolution,
and we show the uncertainty quoted in refs. \cite{Jennings:2012ej, Hahn:2014lca} as shaded regions. Within this relatively large uncertainty,
the results obtained in the viscous description (left panels in figs.\,\ref{fig-3} and \ref{fig-4}) at two loops are again in agreement up to about $k\lesssim 0.2\, h/$Mpc. In addition, the variation when changing
$k_m$ is much smaller than the corresponding cutoff dependence in SPT (right panels in figs.\,\ref{fig-3} and \ref{fig-4}).
The residual $k_m$-dependence yields a quantitative estimate of the theoretical uncertainty.
The difference of the perturbative prediction and the $N$-body results, as well
as the theoretical uncertainty, increase quickly with $k$.
Given that the predictions within the scheme discussed here do not require to fit any free parameters, the level of agreement between the two approaches for $k\lesssim 0.2\, h/$Mpc is remarkable.

\section{Conclusions}

We conclude that the description of dark matter as
a viscous fluid discussed here yields a robust framework for predicting power spectra for $k\lesssim 0.2\, h/$Mpc at $z=0$
without the need to adjust any free parameters. The framework can be 
constructed and extended through a formal analysis 
based on the renormalization group.
We refer the reader to ref. \cite{floerchinger} for details.

%

\begin{thebibliography}{}
%
%

\bibitem{bernardeau0}
  F.~Bernardeau, S.~Colombi, E.~Gaztanaga and R.~Scoccimarro,
  Phys.\ Rept.\  {\bf 367} (2002) 1
  [astro-ph/0112551].



\bibitem{baumann}
  D.~Baumann, A.~Nicolis, L.~Senatore and M.~Zaldarriaga,
  JCAP {\bf 1207} (2012) 051
  [arXiv:1004.2488 [astro-ph.CO]].
  
\bibitem{effective}
  J.~J.~M.~Carrasco, M.~P.~Hertzberg and L.~Senatore,
  JHEP {\bf 1209} (2012) 082
  [arXiv:1206.2926 [astro-ph.CO]].
  
  
\bibitem{Pietroni:2011iz}
  M.~Pietroni, G.~Mangano, N.~Saviano and M.~Viel,
  JCAP {\bf 1201} (2012) 019
  [arXiv:1108.5203 [astro-ph.CO]].

\bibitem{Peebles:1980}
  P.J.E.~Peebles,
  {\itshape The Large-scale Structure of the Universe},
  Princeton University Press (1980).
  
  
\bibitem{blas}
  D.~Blas, S.~Floerchinger, M.~Garny, N.~Tetradis and U.~A.~Wiedemann,
  JCAP {\bf 1511} (2015) 049
  [arXiv:1507.06665 [astro-ph.CO]].
  
  \bibitem{Max1}    
  S.~Matarrese and M.~Pietroni,
    JCAP {\bf 0706} (2007) 026
    [arXiv:astro-ph/0703563].
    
  \bibitem{floerchinger}
    S.~Floerchinger, M.~Garny, N.~Tetradis and U.~A.~Wiedemann,
    arXiv:1607.03453 [astro-ph.CO].
 
\bibitem{nishimichi}
  T.~Nishimichi, F.~Bernardeau and A.~Taruya,
  arXiv:1411.2970 [astro-ph.CO].

\bibitem{Garny:2015oya}
  M.~Garny, T.~Konstandin, R.~A.~Porto and L.~Sagunski,
  JCAP {\bf 1511} (2015) no.11,  032
  doi:10.1088/1475-7516/2015/11/032
  [arXiv:1508.06306 [astro-ph.CO]].
 
 \bibitem{skordis1}
   D.~B.~Thomas, M.~Kopp and C.~Skordis,
   arXiv:1601.05097 [astro-ph.CO].
   
  \bibitem{kunz}
    M.~Kunz, S.~Nesseris and I.~Sawicki,
    Phys.\ Rev.\ D {\bf 94} (2016) no.2,  023510
    [arXiv:1604.05701 [astro-ph.CO]].
   
  \bibitem{skordis2}
    M.~Kopp, C.~Skordis and D.~B.~Thomas,
    arXiv:1605.00649 [astro-ph.CO].
  
      \bibitem{Kim:2011ab}
        J.~Kim, C.~Park, G.~Rossi, S.~M.~Lee and J.~R.~Gott III,
        {\em J.\ Korean Astron.\ Soc.}\  {\bf 44} (2011) 217
        [arXiv:1112.1754 [astro-ph.CO]].

\bibitem{Pueblas:2008uv}
  S.~Pueblas and R.~Scoccimarro,
  Phys.\ Rev.\ D {\bf 80} (2009) 043504
  [arXiv:0809.4606 [astro-ph]].

  
\bibitem{Jennings:2012ej}
  E.~Jennings,
  Mon.\ Not.\ Roy.\ Astron.\ Soc.\  {\bf 427} (2012) L25
  [arXiv:1207.1439 [astro-ph.CO]].

\bibitem{Hahn:2014lca}
  O.~Hahn, R.~E.~Angulo and T.~Abel,
  Mon.\ Not.\ Roy.\ Astron.\ Soc.\  {\bf 454} (2015) 3920
  [arXiv:1404.2280 [astro-ph.CO]].

\end{thebibliography}
%
%

\end{document}